# Stability of temporal solitons in uniform and "managed" quadratic nonlinear media with opposite group-velocity dispersions at fundamental and second harmonics

Peter Y. P. Chen[1] and Boris A. Malomed[2]

[1]School of Mechanical and Manufacturing Engineering,
University of New South Wales, Sydney 2052, Australia

[2]Department of Physical Electronics, School of Electrical Engineering,
Faculty of Engineering, Tel Aviv University
Tel Aviv 69979, Israel

Abstract

The problem of the stability of solitons in second-harmonic-generating media with normal group-velocity dispersion (GVD) in the second-harmonic (SH) field, which is generic to available $\chi^{(2)}$ materials, is revisited. Using an iterative numerical scheme to construct stationary soliton solutions, and direct simulations to test their stability, we identify a full soliton-stability range in the space of the system's parameters, including the coefficient of the group-velocity-mismatch (GVM). The soliton stability is limited by an abrupt onset of growth of tails in the SH component, the relevant stability region being defined as that in which the energy loss to the tail generation is negligible under experimentally relevant conditions. We demonstrate that the stability domain can be readily expanded with the help of two "management" techniques (spatially periodic compensation of destabilizing effects) — the dispersion management (DM) and GVM management. In comparison with their counterparts in optical fibers, DM solitons in the $\chi^{(2)}$ medium feature very weak intrinsic oscillations.

## I. Introduction

Studies of optical solitons in second-harmonic-generating media (those featuring the $\chi^{(2)}$ nonlinearly) are a vast research area. Many experimental and theoretical aspects of the soliton dynamics solitons in this setting have been understood well enough, as reviewed in Refs. [1]. Most experiments with $\chi^{(2)}$ solitons were performed in the spatial domain, where they represent planar or cylindrical beams composed of mutually locked fundamental-frequency (FF) and second-harmonic (SH) waves [2]. Experiments with $\chi^{(2)}$ solitons in the temporal domain are more difficult, due to the requirements of having anomalous group-velocity dispersion (GVD) at both harmonics, and securing the group-velocity synchronism between them. Essential is not only the sign of the GVD, but also its size – it must be strong enough for the respective dispersion length to be less than ~ 1 cm, as the size of experimental samples is limited to a few centimeters. Strong GVD also provides a possibility to compensate the group-velocity mismatch (GVM) between the FF and the SH components, and control the soliton's *walk* [3]. In available $\chi^{(2)}$ optical materials, sufficiently large anomalous GVD at the FF is accessible, without significant linear absorption, if the FF is chosen in the infrared. However, the corresponding SH usually falls into a wavelength interval where the GVD ranges from nearly zero to large normal values.

An experimental technique that allows one to emulate the missing strong anomalous GVD uses tilted wave fronts. This method has made it possible to create temporal $\chi^{(2)}$ solitons [4], as well as spatiotemporal ones, which are self-trapped in the temporal and one transverse directions [5], while the other transverse direction was sacrificed to create the tilted fronts. Prior to the experiment, spatiotemporal solitons in $\chi^{(2)}$ media were predicted in Refs. [6] (see also review [7]). In addition to their fundamental significance, temporal and spatiotemporal solitons are of interest to potential applications, as they may provide switching rates in the THz range, and other features promising to the implementation of all-optical data-processing schemes [8]. It is also worthy to note that the use of $\chi^{(2)}$ effects, in a combination with the Kerr nonlinearity, makes it possible to experimentally demonstrate a novel type of the evolution of temporally localized optical pulses, namely, self-steepening without self-phase modulation [9].

Although soliton solutions do not exist in the system of coupled-mode equations for the second-harmonic generation in the dispersive medium with normal GVD at the SH, it was demonstrated, by means of the variational approximation and in a numerical form, that stable solitons, both spatiotemporal [10] and temporal [11] ones, with tiny (virtually invisible) delocalized "tails" do exist in this case, in a limited range of parameters. In physical terms, such "quasi-solitons" do not differ from their counterparts described by rigorously localized solutions.

It is relevant to mention that the system of coupled-mode equations for the second-harmonic generation in the temporal domain demonstrates the widest variety of (quasi-) solitary-wave solutions in the case of anomalous GVD at the FF and normal GVD at the SH [12] (although many of those solutions may be unstable). As said above, a crucially important issue for the physical implementation, as well as for potential applications, is to identify a parameter region where "tails", inevitably attached to the soliton's "body" in this case, may be effectively suppressed (we here consider only single-peak solitons, i.e., *fundamental* ones). In other words, in this region the generation of the tails will not cause instability of temporal solitons. These issues are the subject of the present work.

The paper is organized as follows. In Section II, we introduce the model and outline numerical methods used to construct numerically exact steady-state solutions, whose stability will then be tested. In Section III, we focus on the analysis of the tail formation and soliton stability in the system with zero GVM. Results of the stability investigation are summarized in the form of a relevant chart in the system's parameter space. Outside the stability domain, solitons lose energy to the generation of tails, which eventually leads to destruction of the solitons.

Well-known results from fiber and bulk optics with the cubic nonlinearity [13] suggest that the technique of the dispersion management (DM) may help to increase the range of the normal GVD at the SH that admits stable transmission of the temporal solitons, without the generation of conspicuous tails. In this connection, it is relevant to mention that the application of the "management" (periodic alternation) to the local wave-vector mismatch of second-harmonic generating systems exerts a stabilizing effect on spatial $\chi^{(2)}$ solitons, as shown in recent work [14]. In Section IV, we demonstrate that the stability limits for temporal solitons may indeed be substantially extended by means of the DM. Effects of the DM on the pulse shape and its transmission characteristics are illustrated by means of generic examples. The suppression of the tail generation by means of the DM is also illustrated (in Section IV) by means of a crude analytical approximation.

In Section V, we include the GVM into the present model. After confirming results presented in Ref. [11], according to which only a small size of the constant GVM coefficient admits steady transmission of the solitons, we present a new model, based on *GVM management*, which is suggested by the DM scheme, as well as by the *tandem scheme* [15 - 17]. It is demonstrated that large GVM may be tolerated if this technique is used. The paper is concluded by Section VI, where we also briefly discuss conditions for the experimental realization of the proposed settings.

**II.   The model and numerical methods**

The system of coupled-mode equations governing the co-transmission of the FF and SH waves in the dispersive $\chi^{(2)}$ medium is well known. In the normalized form, it can be written as [1,3-5,11]:

$$iu_z + \frac{1}{2}D_1 u_{\tau\tau} + vu^* = 0. \qquad (1)$$

$$2i(v_z + cv_\tau) + \frac{1}{2}D_2 v_{\tau\tau} - qv + \frac{1}{2}u^2 = 0. \qquad (2)$$

Here, $-D_1$ and $-D_2$ are the GVD coefficients at the FF and the SH, respectively (i.e., positive coefficients $-D_{1,2}$ correspond to the anomalous dispersion), $c$ is the GVM parameter, $q$ represents the wave-vector mismatch between the FF and SH waves, and the asterisk stands for the complex conjugation.

Equations (1) and (2) conserve the momentum, Hamiltonian, and the *Manley-Rowe invariant*, which is frequently called energy, in terms of the temporal-domain optics,

$$E \equiv E_u + E_v = \int_{-\infty}^{+\infty} [|u(\tau)|^2 + 4|v(\tau)|^2] d\tau. \tag{3}$$

In our numerical simulations, the integration in this expression will be limited to the actual computational window. It can be checked that the effect of this limitation on accuracy of the results is completely negligible.

For $c = 0$, $q < 0$, and the GVD coefficients satisfying conditions $0 < D_2 < 4D_1$, Eqs. (1), (2) have a well-known exact solution for temporal solitons [18],

$$u = \frac{A e^{ikz}}{\cosh^2(a\tau)}, \qquad v = \frac{B e^{2ikz}}{\cosh^2(a\tau)}, \tag{4}$$

$$k = \frac{D_1 q}{D_2 - 4D_1}, \quad a^2 = \frac{q}{2(D_2 - 4D_1)}, \quad A^2 = \frac{9 D_1 D_2 q^2}{2(D_2 - 4D_1)^2}, \quad B = \frac{3 D_1 q}{2(D_2 - 4D_1)}. \tag{5}$$

General soliton solutions, in the form of $u(z,\tau) = e^{ikz}U(\tau)$, $v(z,\tau) = e^{2ikz}V(\tau)$ with arbitrary positive propagation constant $k$, can be found in a numerical form, or described by means of the variational approximation [19].

For $c = 0$, the stationary version of Eqs. (1) and (2) takes the form of the boundary-value problem, with the fields that should vanish (if this is possible) at $|\tau| \to \infty$:

$$kU - \frac{D_1}{2} \frac{d^2 U}{d\tau^2} - VU^* = 0.$$
$$4kV - \frac{D_2}{2} \frac{d^2 V}{d\tau^2} + qV - \frac{1}{2}U^2 = 0. \tag{6}$$

To solve this problem, we used a finite-element-spectral method, that reduces Eqs. (6) to a set of nonlinear algebraic equations [20]. The computations were performed in a domain sufficiently large to make the results insensitive to the size of the domain, with zero boundary conditions at the edges. To get a formally closed-form system, Eq. (3) was added to Eqs. (6), treating $k$ as another unknown. Then, the standard Newton's iterative procedure could be used to solve the resultant set of algebraic equations.

In many previous works, Eqs. (6) were solved with normalization $k = 1$, see, e.g., Ref. [11]. We here choose another normalization, looking for (quasi-) soliton solutions subject to condition $E = 20$ [recall $E$ is the total energy defined by Eq. (3)]. Actually, the latter normalization is more physically relevant, as the energy, unlike the wavenumber, is a directly observable physical characteristic.

Below, we will also consider the DM version of Eqs. (1), (2), with the dispersion coefficients made periodic functions of $z$. For this case, we have found quasi-stationary (regularly oscillating) solitary-wave solutions, by means of a properly adjusted iterative method that was previously elaborated for the search of stable solitons in the DM model with the cubic nonlinearity [21] (if a stable DM soliton does not exist, the iterative algorithm does not converge). Further, if the GVM term is present in Eq. (2), i.e., $c \neq 0$, we were able to use a variant of the iterative method, modified similar to how it was done in Ref. [22] for the analysis of effects of the polarization-mode dispersion on DM solitons.

In addition to the DM scheme, we will analyze a model of the "GVM management", which is built as a periodic chain of cells with opposite signs of the GVM, making the average value of the GVM coefficient equal to zero. As said above, a $\chi^{(2)}$ model utilizing the mismatch management [14] was studied recently, but, to the best of our knowledge, a similar scheme was not considered to effectively cancel the GVM, although the earlier proposed "tandem scheme" [15,16], based on the periodic alternation of $\chi^{(2)}$ and linear segments, pursued a somewhat similar idea. Similar to the case of the DM, the tandem setting was predicted to stabilize *guiding-center* (strongly pulsating) solitons [15]. The efficiency of the tandem scheme was demonstrated experimentally [23], and its generalization was also elaborated for the compensation of the GVM affecting spatiotemporal (multidimensional) solitons [17]

After stationary (or quasi-stationary, in the DM model) soliton solutions were found, we have tested their stability in long simulations of Eqs. (1), (2), using the split-step algorithm. The typical transmission distance in the simulations was $z = 80$, which is quite sufficient to distinguish between stable and unstable solitons, see Fig. 7 below (in all cases, this distance exceeds 10 soliton's dispersion lengths, and it has been checked, in typical cases, that the character of the soliton's stability does not alter in much longer simulation runs). As concerns DM solitons, the convergence of the iterative method for them is not sufficient to conclude if they are stable, as this method actually involves a short propagation distance, therefore the stability of the DM solitons was also tested in long simulations.

In fact, the main issue concerning the soliton's stability is not a possibility of the destruction of the soliton's "body" by small perturbations (it was easy to verify that all the solitons are stable in that respect), but, as said above, the effective destabilization through the energy loss to the generation of tails in the SH field, in the case of $D_2 < 0$. An adopted technical stability criterion was that, at the end of the simulation, the body of the pulse must not lose more than 5% of the initial energy in the SH field. The corresponding soliton will seem completely stable in any possible experimental setting.

For the advance of the simulations along $z$, the computations were carried out, in most cases, with a small step of $\Delta z = 0.001$, which secured high accuracy of the split-step simulations. If the growth of the tails was conspicuous, $\Delta z$ was further reduced to ensure that the growth rate did not depend on $\Delta z$.

**III. Soliton stability in the absence of the group-velocity mismatch**

As said above, we have identified stability borders of a family of stationary (quasi-) soliton solutions to Eqs. (6), in the plane of parameters $D_2 < 0$ and $q$, by means of systematic direct simulations of Eqs. (1) and (2). We present a summary of these results in the form of the stability diagram in Fig. 1, fixing the normalization to $D_1 \equiv 2$ (recall that the GVD at the FF is always anomalous, in situations relevant to the study of $\chi^{(2)}$ solitons) and, as mentioned above, $E = 20$. The rightmost stability border, corresponding to the model with constant GVD coefficients, complies with results reported in Ref. [11], while the borders for the DM model (which are discussed in detail below) represent essentially new

result

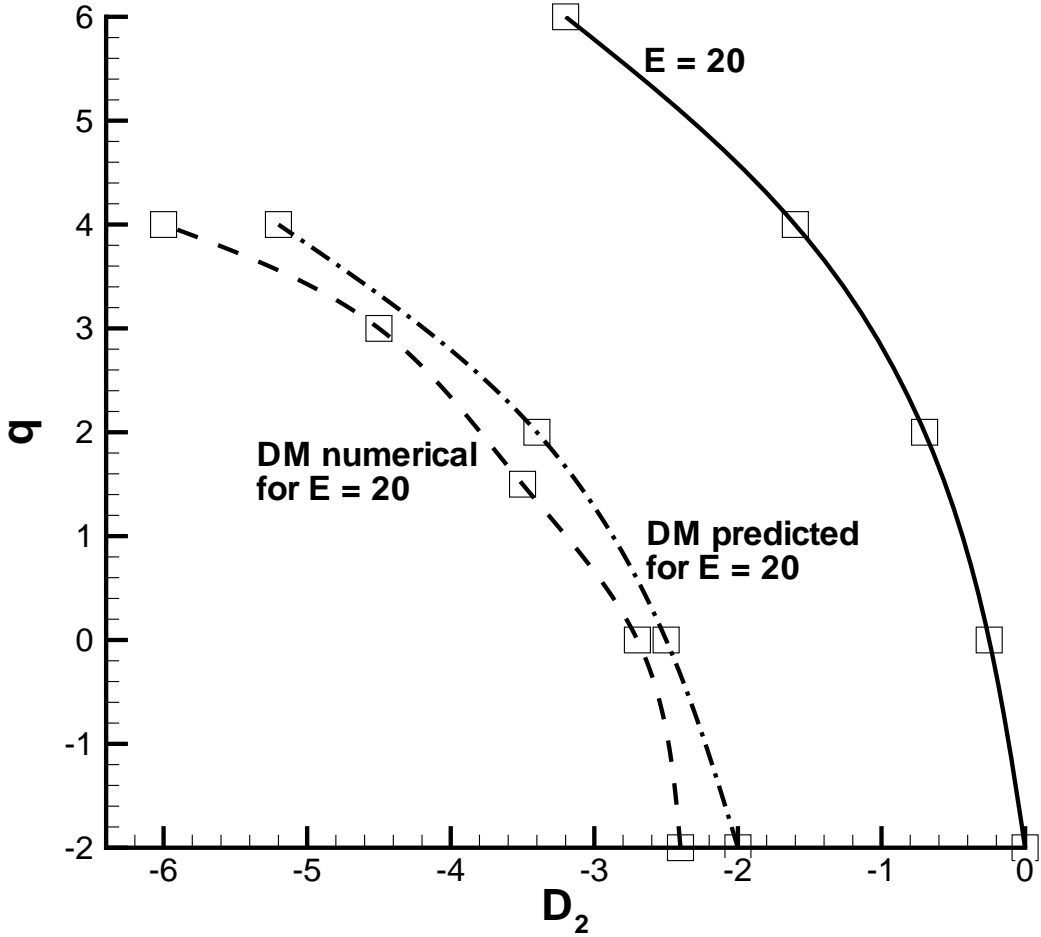

**Figure 1.** Stability borders for temporal solitons in the system without GVM ($c = 0$), and normalization $D_1 \equiv 2$. Solitons are stable to the right of the curves corresponding to fixed energy $E = 20$. The rightmost border corresponds to the system with constant GVD coefficients, while two other borders are obtained in the model with the dispersion management (see Section IV).

A noteworthy feature of the diagram is that the stability border can be pushed farther into the region of the normal second-harmonic GVD by increasing mismatch parameter $q$. This finding is not surprising, as, for large $|q|$, Eq. (2) yields $v \approx u^2/(2q)$ in the well-known cascading approximation. The substitution of this relation in Eq. (1) leads to the usual cubic nonlinear Schrödinger (NLS) equation for the FF field, $iu_z + u_{\tau\tau} + (2q)^{-1}|u|^2 u = 0$ (recall that $D_1 \equiv 2$). Obviously, this equation gives rise to solitons for $q > 0$, and it does not admit bright solitons for $q < 0$, in agreement with Fig. 1.

Figures 2 and 3, where we fix $q = 2$, demonstrate how the growth of the tails and the onset of the effective instability of the solitons in the SH component are associated with the increase of the normal GVD coefficient, $-D_2$. Well inside the stability region (but at $D_2 < 0$), the solutions feature practically no tails, while the tail's amplitude drastically increases across the instability boundary. In Fig. 3, the same set of pulses is shown after they have passed the distance of $z = 40$. Comparison with Fig. 2 demonstrates that, inside the stability region, *viz.*, for $D_2 = 0$ and $D_2 = -0.3$, the propagation gives rise to practically no change in the shape of the solitons, including the absence of tails. For $D_2 = -0.7$, which is identified as the stability limit, the tails have started to grow, absorbing the energy from the central portion of the pulse. The energy loss is, however, smaller than the limit of 5% set above. Beyond the stability border, the loss of the energy exceeds the limit.

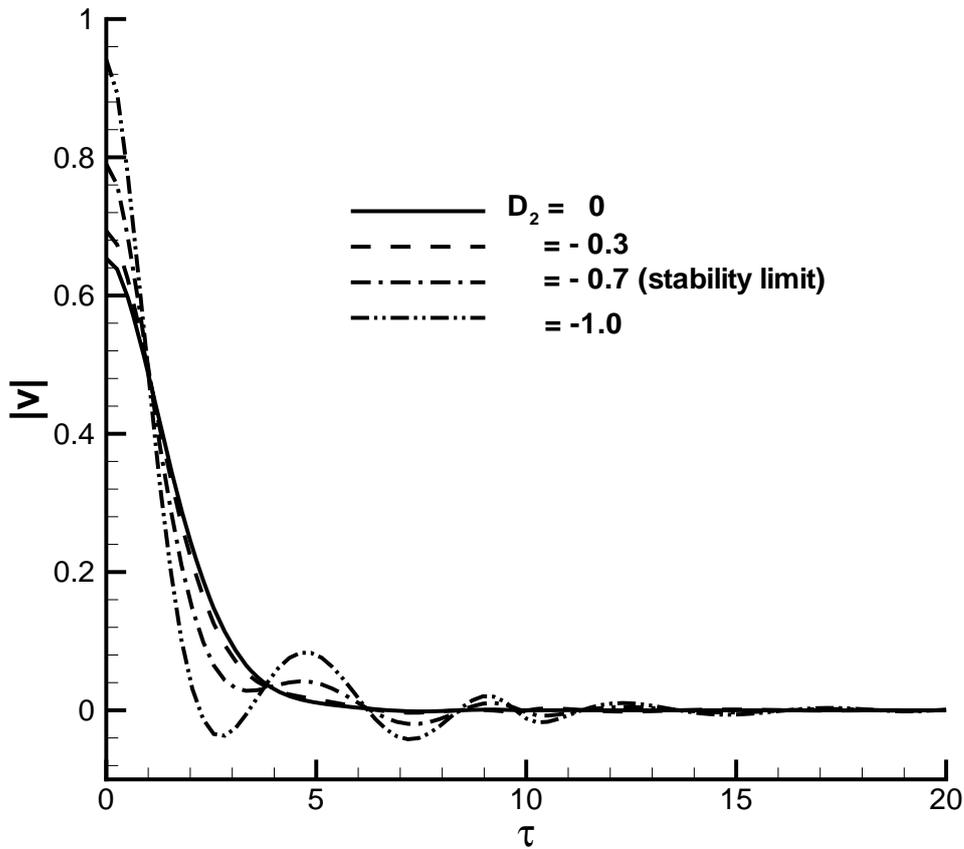

**Figure 2**. The shape of the second-harmonic component of stationary (quasi-) soliton solutions for $D_2 = 2, E = 20,$ and $q = 2$. The development of tails with the increase of $|D_2|$ is obvious.

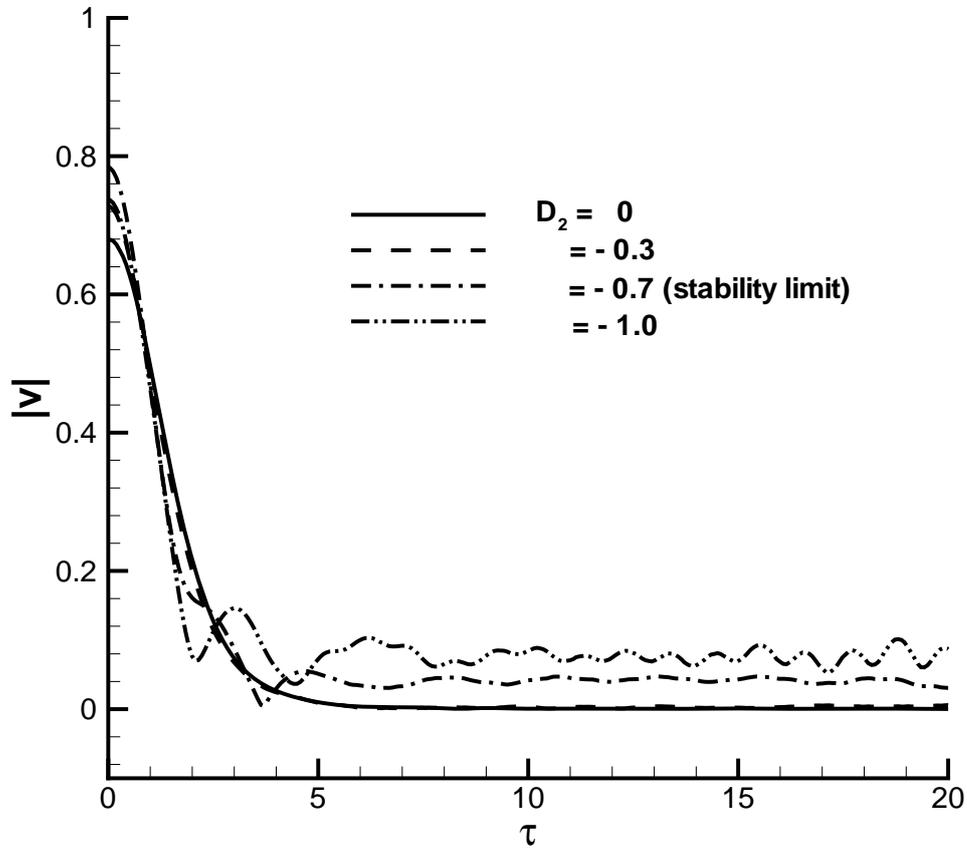

**Figure 3**. The shapes of the second-harmonic components in the set of solitary waves from Fig. 2 after passing distance $z = 40$.

In addition to Figs. 1 and 3, the growth of the share of the total energy in the tails with the increase of $-D_2$ is shown, for fixed mismatch $q$, in Fig. 4. This figure demonstrates the steep decrease of the energy loss as the system shifts into the stability region. Together with Fig. 4, it is relevant to display the share of the total energy of the soliton in its SH component. Figure 5 displays this share calculated along the stability border, which was identified in Fig. 1 (in accordance with the definition of the stability border, the part of the SH energy transferred to its tails is still negligible, at these points). In accordance with the above-mentioned cascading approximation, the SH energy share sharply drops with the increase of mismatch. Additionally, Fig. 6 shows the same energy share as a function of $D_2$, for fixed mismatch, $q = 4$. All points belonging to the latter plot are taken at the border of or inside the stability region.

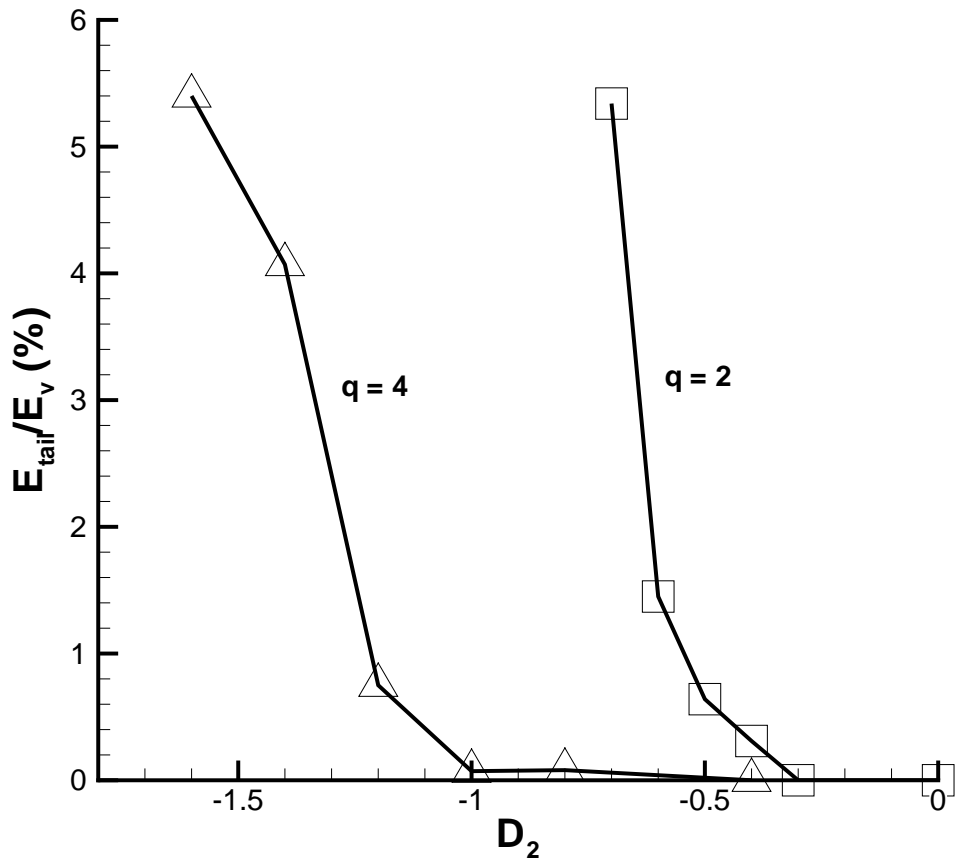

**Figure 4**. The share of the total energy in the tails of the solitary waves which have passed the distance of $z = 40$, for different fixed values of mismatch $q$. The highest point of each plot corresponds to the stability limit, as defined above.

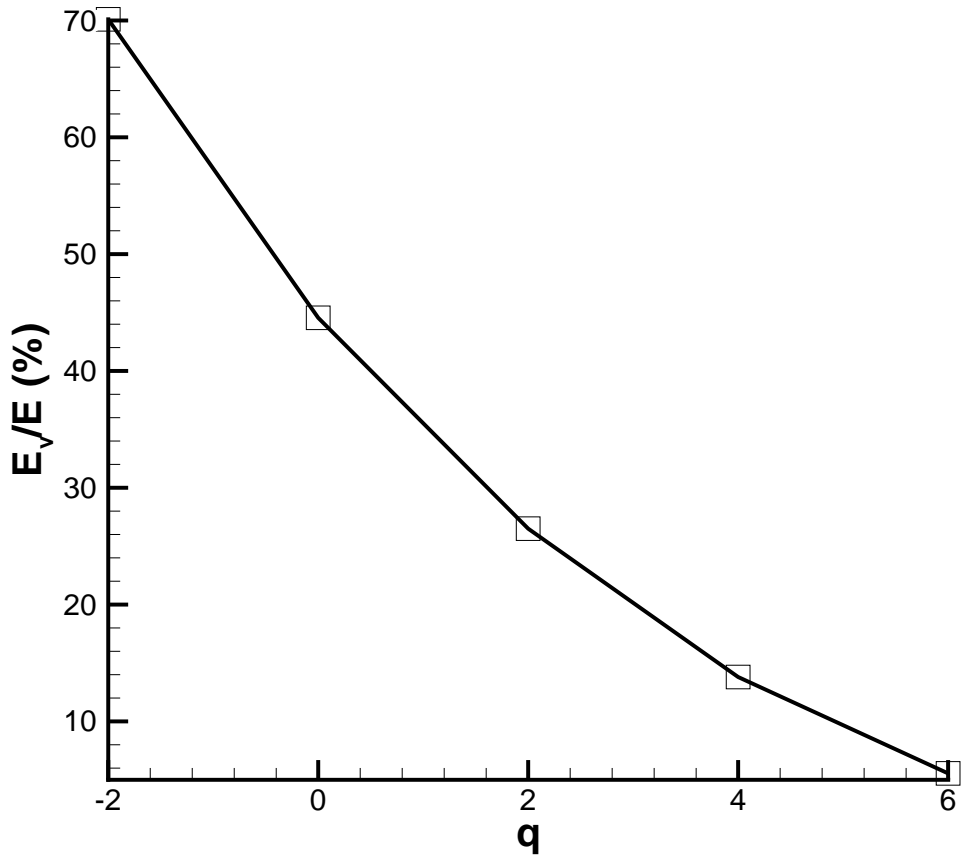

**Figure 5**. The share of the total soliton's energy, $E = 20$, in the second-harmonic component, along the border of the stability region, which was identified in Fig. 1.

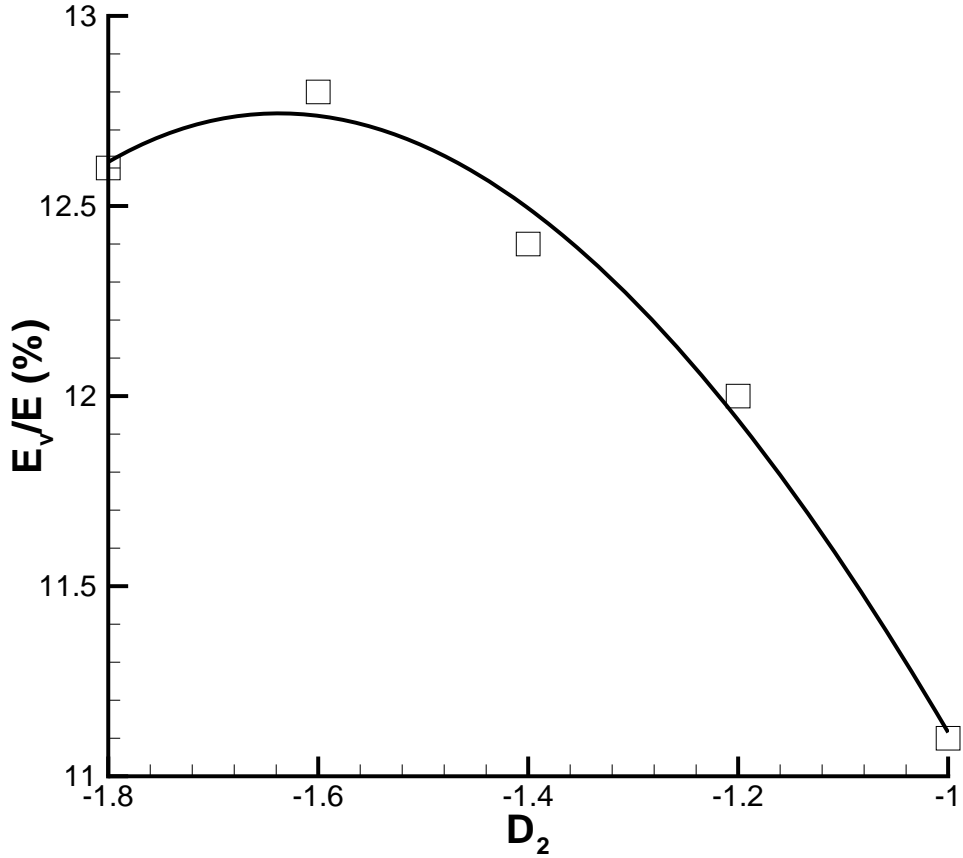

**Figure 6.** The same energy share as in Fig. 5 versus the second-harmonic GVD coefficient (in the region of the normal dispersion), for a fixed (typical) value of the mismatch, $q = 4$.

In the experiment, the soliton cannot be prepared with the exact shape prescribed by the solution to Eqs. (6). Therefore, it is relevant to additionally test the robustness of stable solitons by launching them at "wrong" values of parameters. A typical example of such a test is displayed in Fig. 7, where we take the soliton obtained as a solution to Eqs. (6) at $E = 20, q = 2, D_1 = 2,$ and $D_1 = -0.7$ (this point falls onto the stability border in Fig. 1), and simulate its propagation at these values of the system's parameters [in Fig. 7(b)], as well as at "wrong" values, namely, for $q = 2.5$ (inside the stability region) and $q = 1$ (beyond the stability border), as shown in Figs. 7(a) and 7(c), respectively. It is seen that, in the former case, the pulse features persistent intrinsic oscillations, due to the fact that it is not adjusted to the respective parameter values, but without development of any instability. On the other hand, in the latter case the pulse quickly decays through intensive generation of tails.

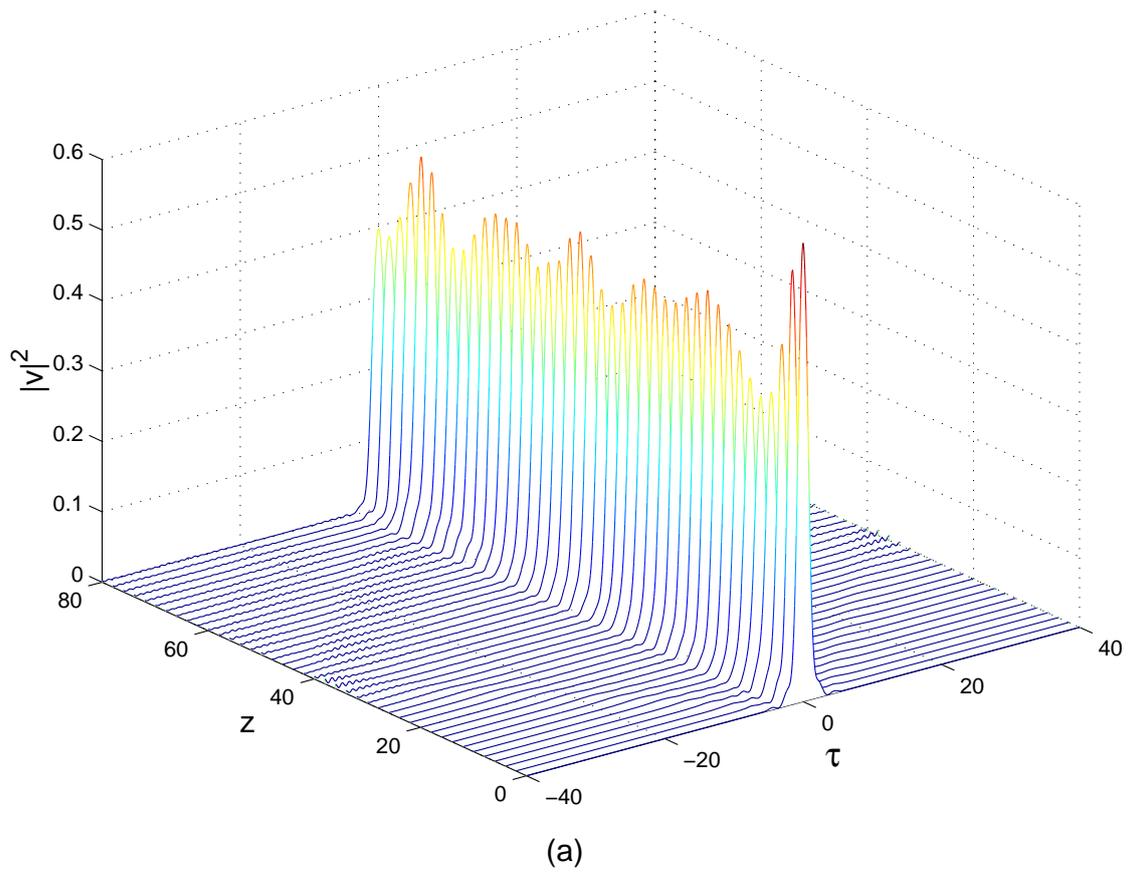

(a)

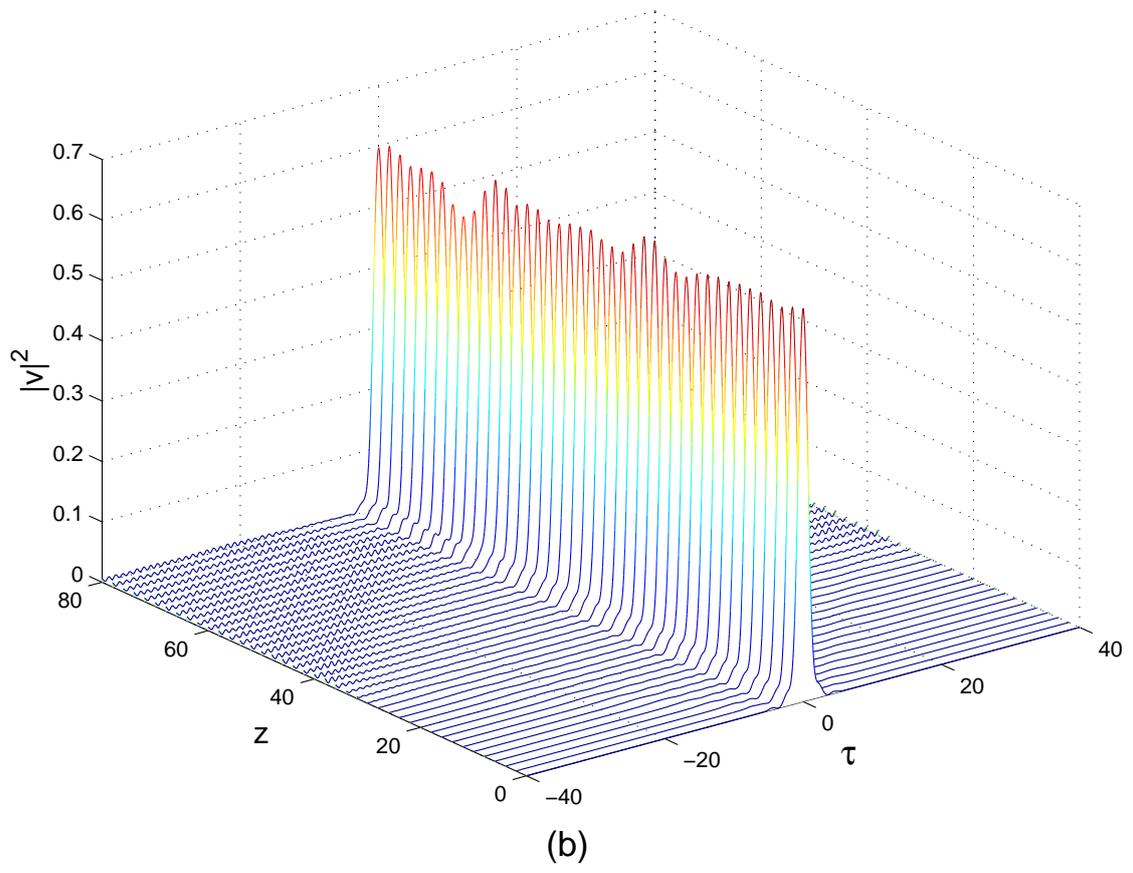

(b)

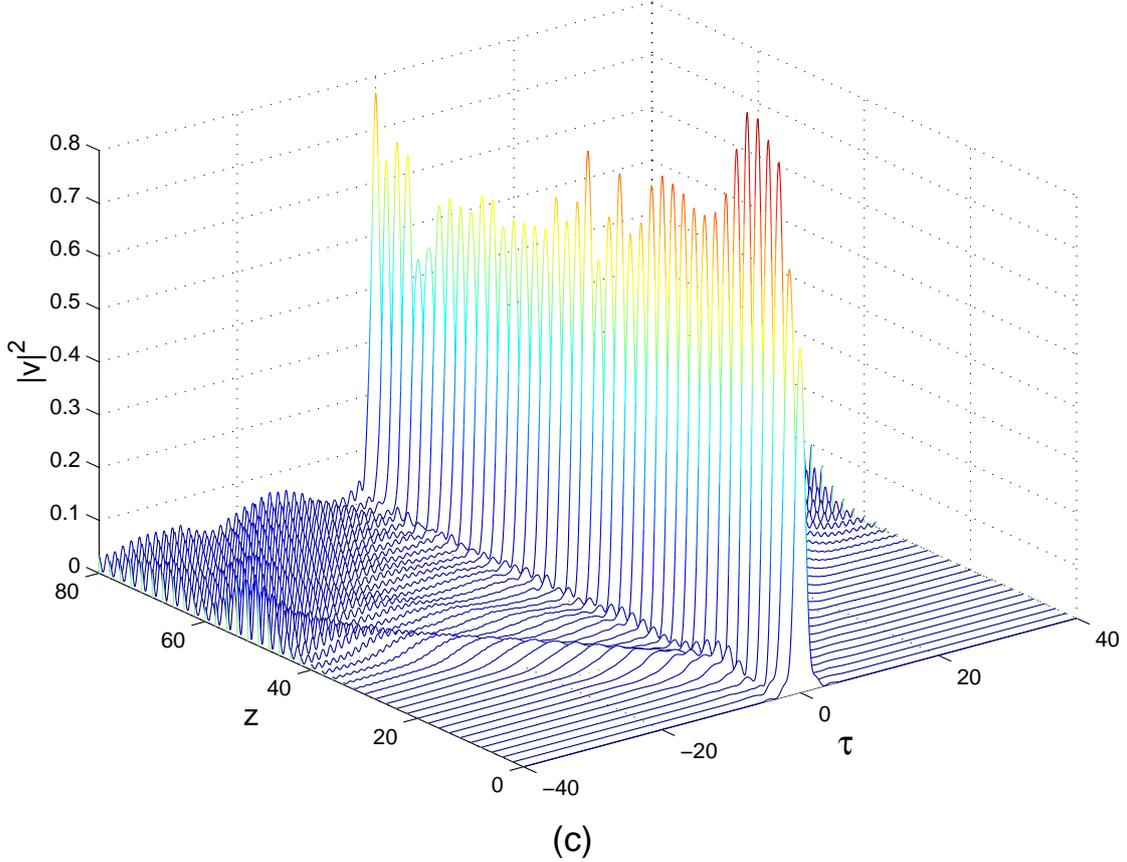

(c)

**Figure 7**. (Color online) The propagation of the pulse prepared for parameter values $E = 20, q = 2, D_1 = 2,$ and $D_1 = -0.7$ (this point belongs to the stability border in Fig. 1) and launched into the system as follows: (a) with a "wrong" value of the mismatch, $q = 2.5$, which falls into the stability region in Fig. 1; (b) with the "correct" mismatch, $q = 2$; (c) with a "wrong" mismatch, $q = 1$, that belongs to the instability region.

## IV. Dispersion management

### A. Numerical results

The DM scheme, results for which are included in the stability diagram displayed in Fig. 1, is defined in terms of the respective *map*: the SH GVD coefficient takes values $D_2 < 0$ and $\widetilde{D}_2 > 0$ in segments of lengths $l$ and $\tilde{l}$, respectively. This arrangement is repeated periodically, with period $l + \tilde{l}$, the corresponding *path-average dispersion* (PAD) coefficient being

$$\langle D_2 \rangle = \frac{lD_2 + \tilde{l}\widetilde{D}_2}{l + \tilde{l}} \qquad (7)$$

(the map does not include a difference in the values of the FF GVD coefficient, $D_1$, between the two segments, but it can be readily included, if necessary for an accurate description of the experiment). For actual iterations of the soliton transmission in the DM system, the DM cell was defined as a "sandwich" composed of three segments, with lengths $l/2 = 1/4, \tilde{l} = 1/2, l/2 = 1/4$, the DM period being $l + \tilde{l} = 1$. In the middle section, the GVD coefficient was set to be $\widetilde{D}_2 = 2$, while in the outer sections the GVD coefficient is negative (and subject to the variation), corresponding to the normal dispersion.

First, we have used the iterative method, as elaborated in Ref. [18], to generate numerically exact solutions for DM solitons in the $\chi^{(2)}$ model. Then, simulating the transmission of these solitons, we could test their stability. The dashed curve in Fig. 1 shows the so identified stability border. The dotted curve, which runs quite close to the dashed one, represents the stability border which could be predicted, in a simple way, by equating the PAD value given by Eq. (7) to that which was found numerically as the stability border in the system with constant coefficients (the solid curve in Fig. 1).

Comparing the results with those obtained in the system with constant GVD, we conclude that the DM makes it possible to considerably extend the stability region. Within the extended region, the tail formation and growth follow exactly the same pattern as in the system without the DM. In particular, the increase of the energy share in the tail, as a function of $D_2$, for the DM systems is shown in Fig. 8. The comparison with Fig. 4 demonstrates that the dependence retains the same character as in the system with constant GVD, except for the shift to much larger values of $-D_2$.

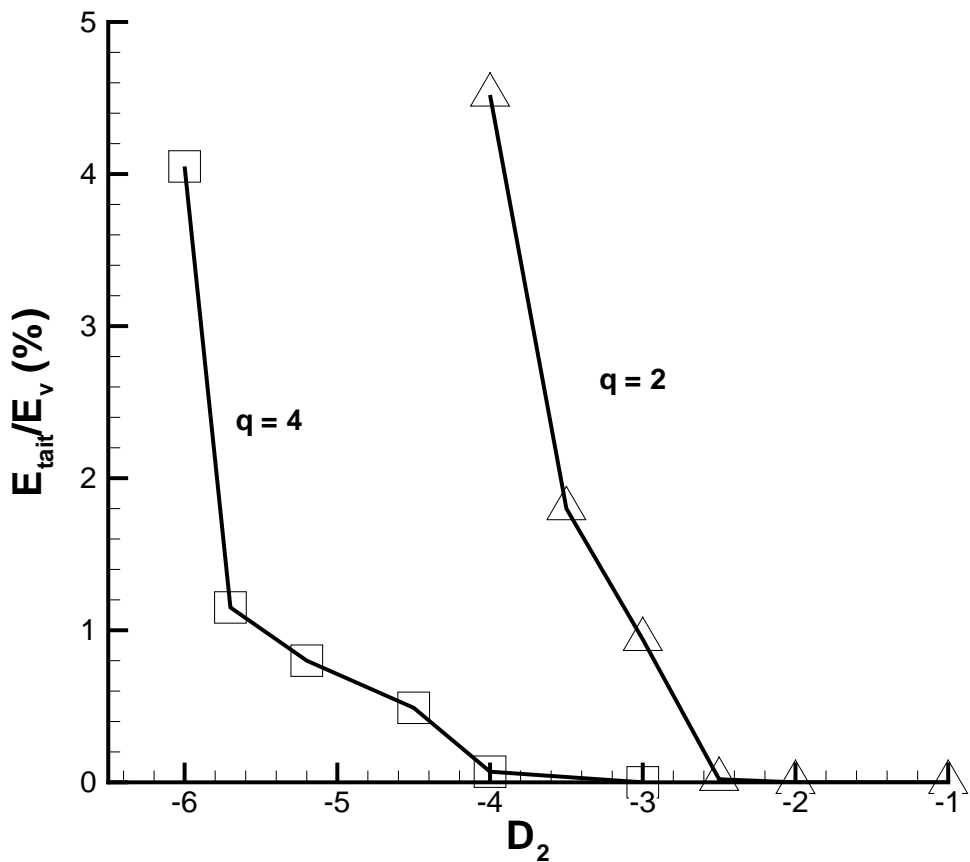

**Figure 8**. The same as in Fig. 4, but in the system additionally stabilized by the dispersion management.

It is worthwhile to notice that, because of the FF-SH mode coupling, the DM solitons exhibit very little "breathing" (periodic shape oscillations) in comparison with the dispersion-managed solitons in the fiber-optic model with the $\chi^{(3)}$ nonlinearity. A typical example demonstrating this feature is displayed in Fig. 9. The DM also produces a little effect on shapes of both the FF and SH components of the solitons, and on the energy distribution between them.

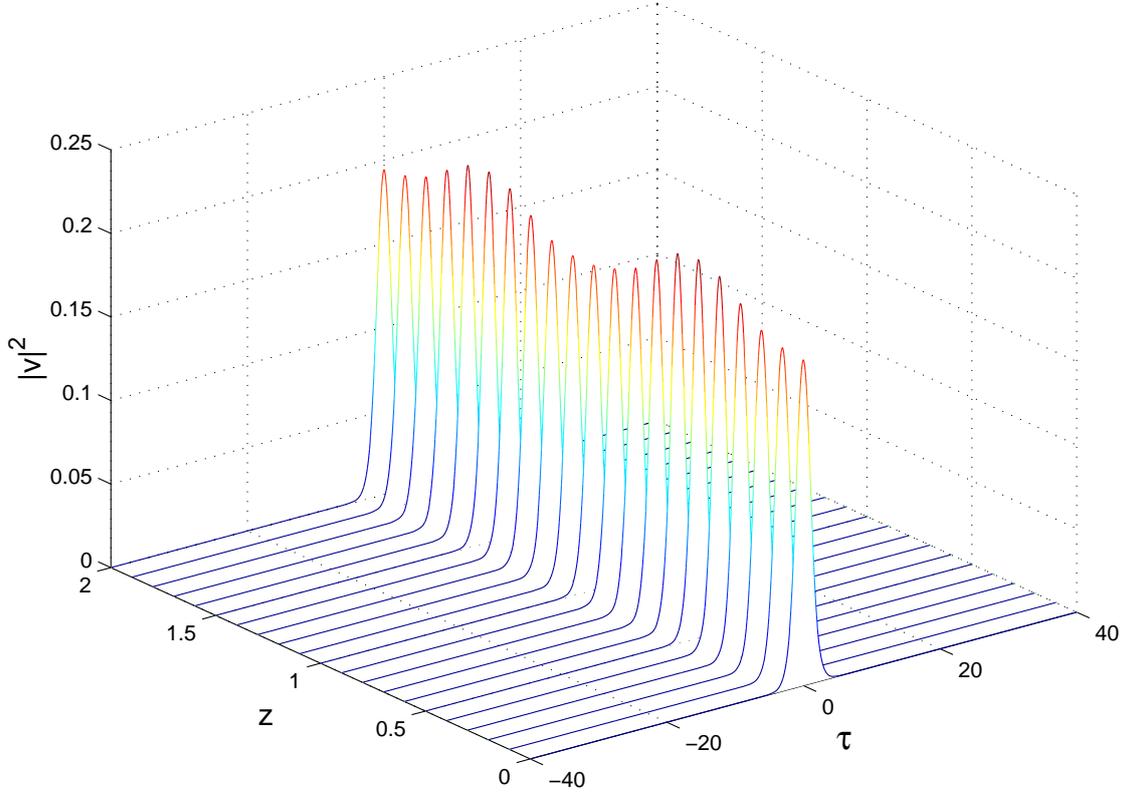

**Figure 9.** (Color online). The evolution of the second-harmonic component in the soliton within two adjacent cells of the dispersion-managed system with $q = 4$ and $D_2 = -4$.

### B. An analytical approach

The suppression of the tail generation by the DM can be illustrated by means of a crude approximation, in which the source of the SH radiation, i.e., term $-(1/2)U^2$ in the second equation in system (6), is approximated by the delta function, assuming that we aim to analyze the SH tail on a temporal scale much larger than the intrinsic width of the soliton:

$$2iv_z + \frac{1}{2}D_2(z)v_{\tau\tau} = \varepsilon\delta(\tau), \tag{8}$$

where $\varepsilon$ is the strength of the effective source, and full matching between the source and the generated field is assumed. The latter condition corresponds to the "most dangerous" regime of the strongest tail generation.

A simple consequence of Eq. (8) is an exact evolution equation for the energy of the SH field component, which is defined as per Eq. (3):

$$\frac{dE_v}{dz} = -4\varepsilon\,\mathrm{Im}\bigl(v(\tau=0)\bigr). \tag{9}$$

Equation (9) can be used to calculate the rate at which the energy is pumped into the soliton's tail in the SH field.

Off the point $\tau = 0$, linear equation (8) has a well-known solution in the form of the chirped Gaussian [13]:

$$v(\tau, z) = \frac{A_0}{\sqrt{1 + \left(i\Delta(z)/2\tau_0^2\right)}} \exp\left\{-\frac{1}{2\tau_0^2} \frac{\tau^2}{\sqrt{1 + \left(i\Delta(z)/2\tau_0^2\right)}}\right\}, \tag{10}$$

where $\Delta(z) \equiv \int_0^z D_2(z)dz$ is the *accumulated dispersion*, while $\tau_0$ and $A_0$ are arbitrary temporal scale and amplitude. The energy of the Gaussian field is

$$E_v = 4\sqrt{\pi} |A_0|^2 \tau_0, \tag{11}$$

see Eq. (3). In the case of the standard piece-wise DM map which is defined above, $\tau_0$ determines the *DM strength* [13],

$$S = 1.443 \frac{\tilde{D}_2 \tilde{l}_2 + |D_2| l_2}{\tau_0^2} \tag{12}$$

(the numerical factor in this definition accounts for the difference between $\tau_0$ and FWHM width of the Gaussian pulse).

The substitution of expression (10) into Eq. (9) yields the following final expression for the rate of pumping the energy into the tail:

$$\frac{dE_v}{dz} = \varepsilon \sqrt{\frac{2E_v}{\sqrt{\pi}\tau_0}} \frac{\sqrt{\sqrt{1 + \Delta^2(z)/(4\tau_0^2)} + 1} - \text{sgn}(\Delta(z))\sqrt{\sqrt{1 + \Delta^2(z)/(4\tau_0^2)} - 1}}{\sqrt{1 + \Delta^2(z)/(4\tau_0^2)}}, \tag{13}$$

where relation (11) was used to express $|A_0|$ in terms of $E_v$. This result is still cumbersome in the general case. However, in the limit case of the *strong DM*, i.e., $S \gg 1$ [13], and assuming zero PAD, $\langle D_2 \rangle = 0$, i.e., full GVD compensation, see Eq. (7) (these are conditions which allow one to explore the effects of the DM in the pure form), the averaging of expression (13) over the DM map yields an asymptotic result,

$$\left\langle \frac{dE_v}{dz} \right\rangle = 4\varepsilon \sqrt{E_v} \left(\frac{1.443}{\pi |D_2| l_2}\right)^{1/4} S^{-1/4}.$$

The purport of the latter result is that, within the framework of the crude analytical model, the rate at which the energy flows from the soliton's body into the SH tail decays $\sim S^{-1/4}$ in the regime of the strong DM.

### V. The systems with the group-velocity mismatch (GVM)

The GVM term in Eqs. (2), which is proportional to coefficient $c$, breaks the symmetry of the soliton's shape with respect to change $\tau \to -\tau$, as shown in Fig. 10 (cf. Fig. 2 for $c = 0$). Which is most essential, the GVM imposes severe limitations of the size of the stability region, or, in other words, the simulations of the soliton propagation make it possible to identify the maximum value of $c$ beyond which the solitons loose their stability. These results are summarized in Fig. 11, which displays the largest value of the GVM coefficient up to which the soliton remains stable, as a function of the mismatch parameter, $q$, for two cases: along the stability border in Fig. 1, and at a set of arbitrarily chosen points inside the stability region.

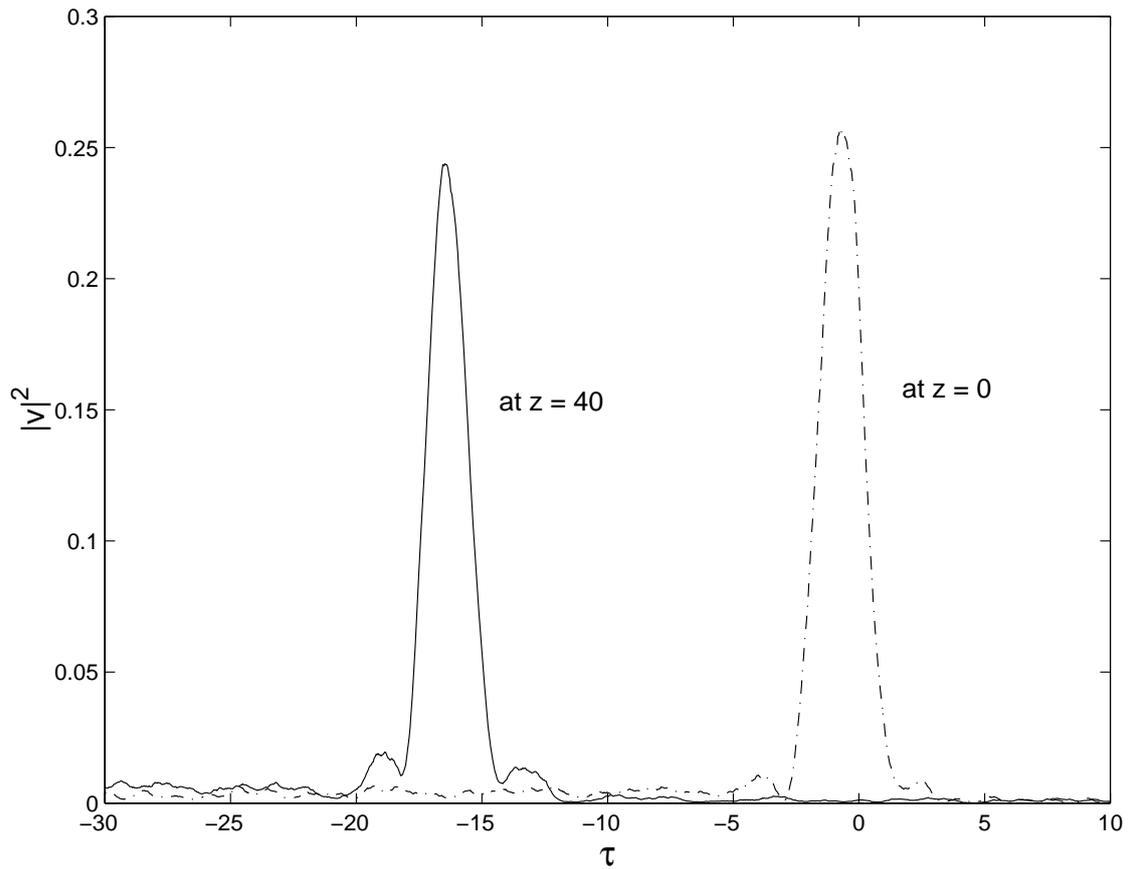

**Figure 10**. Profiles of asymmetric solitons obtained by means of the iterative method from the stationary version of Eqs. (1) and (2) with $c=1$.

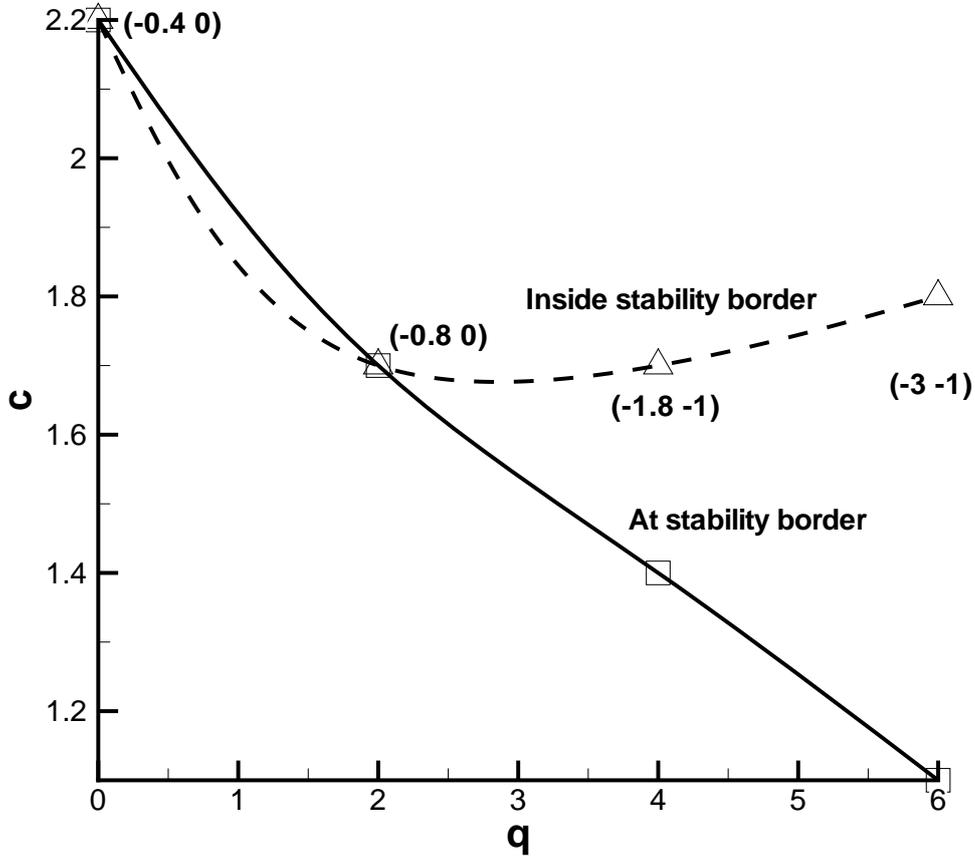

**Figure 11**. The largest value of the group-velocity-mismatch parameter, $c$, beyond which the soliton loses its stability. Pairs of numbers in the brackets are values of $D_2$ for pairs of points appertaining to common values of the wave-vector mismatch coefficient ($q = 0, 2, 4, 6$) and taken, respectively, at the stability border and inside the stability region.

To overcome the restriction imposed on the size of $c$, it is possible to develop a "management" scheme similar to the DM, *viz.*, to consider a system built as a periodic alternation of segments with opposite values of $c$, so as to make the respective average value equal to zero. We have explored this possibility, carrying out systematic simulations of the propagation of solitons in the GVM-managed model. The system was built as the periodic combination of segments of lengths $l_+ = l_- = 1$ that feature opposite values of the GVM coefficient, hence its path-average value is zero, cf. Eq. (7). Again, the iterative method was used to generate appropriate pulses, whose transmission was then explored.

As expected, the GVM-management scheme readily provides for the stabilization of solitons. A generic example is shown, for $E = 20, D_1 = 2, D_2 = -0.6$ and $q = 3$, in Fig. 12. The soliton propagation is seen to be completely stable, without any tangible generation of tails, even for very large local values of the GVM parameter, $c_+ = -c_- = 14$ (cf. Fig. 11, where, in the system with constant $c$, stable propagation is limited to $|c| \leq 2$). A noticeable change in the shape of the SH component of the soliton, caused by the GVM management, is an increase of its width, as seen in Fig. 13.

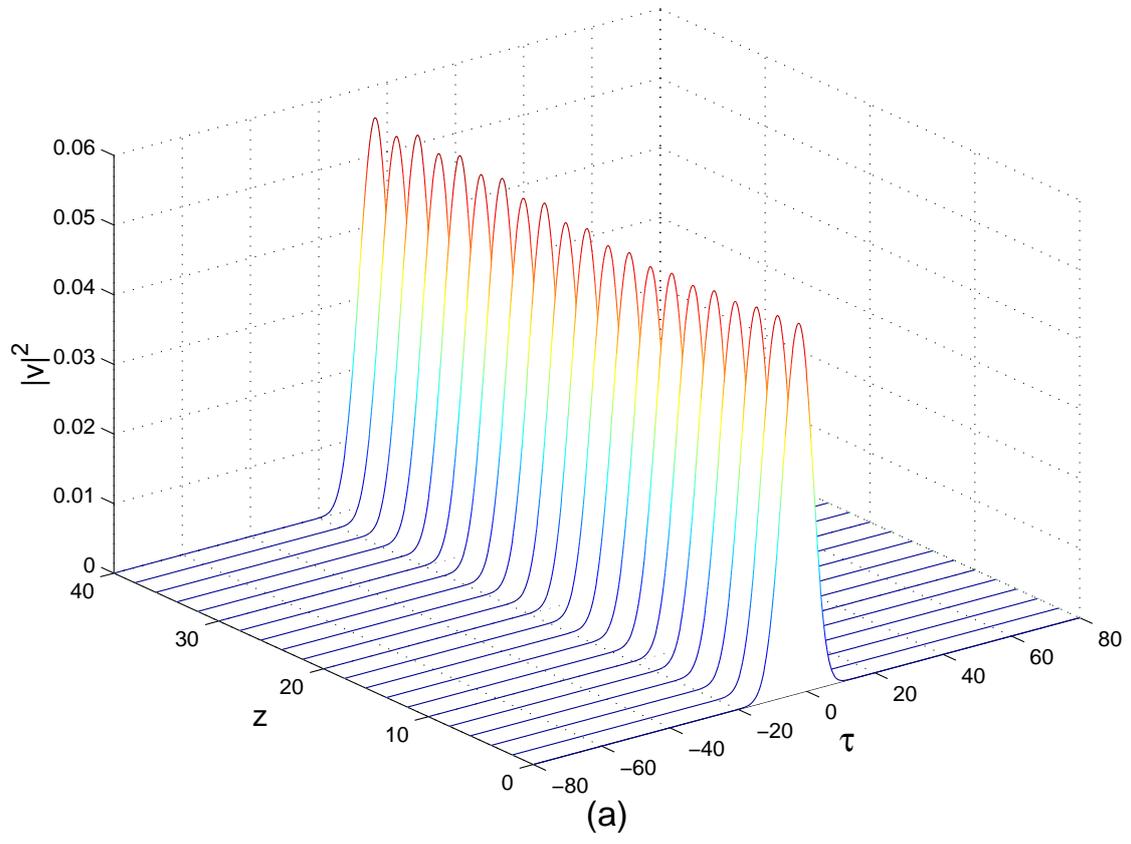
(a)

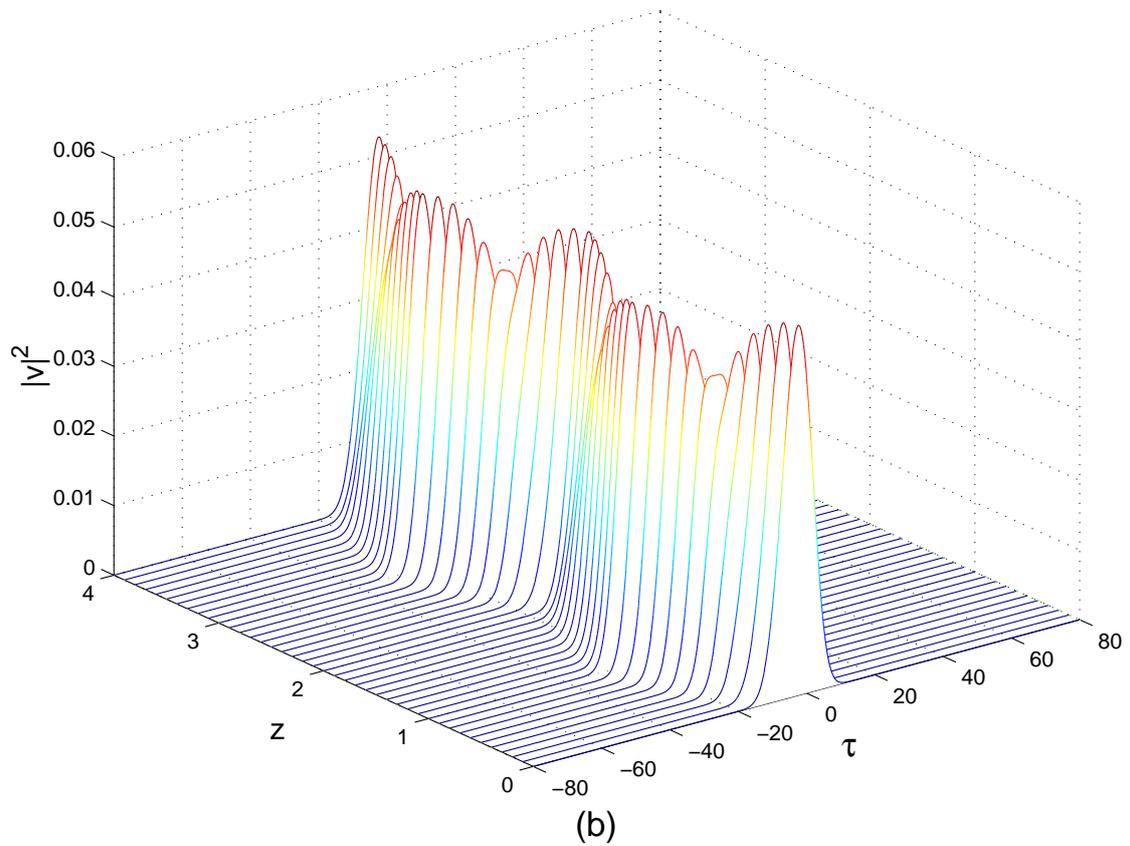

**Figure 12**. (Color online). (a) The long-distance propagation of a typical soliton in the system with the strong GVM management, corresponding to $|c| = 14$ (only the SH component is shown). (b) A detailed picture of the evolution of the pulse within two adjacent cells of the GVM-management scheme.

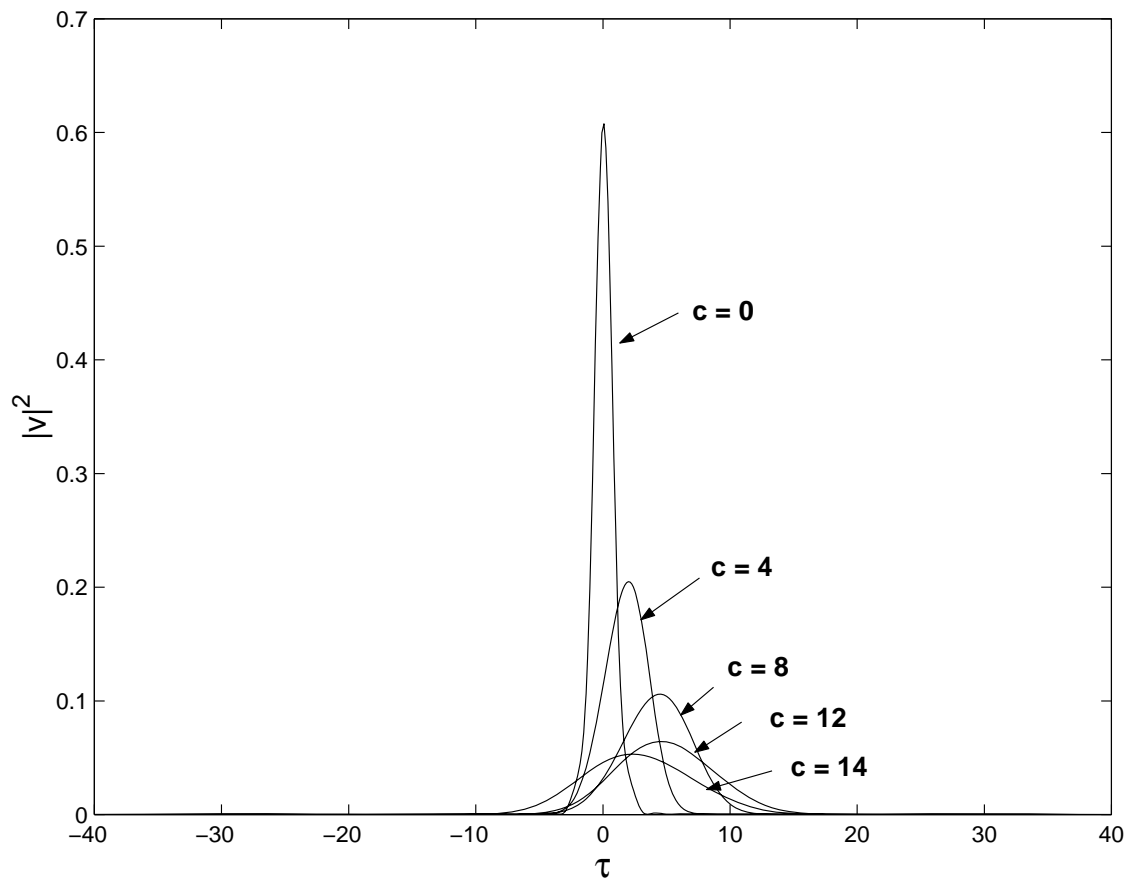

**Figure 13**. The broadening of the second-harmonic component of stable pulses in the system with the group-velocity-mismatch management, with the increase of the absolute value of the CVM parameter, $c$. The pulse shapes are displayed at the transmission stage when they are narrowest.

## VI. Conclusions

We have revisited the problem of the existence of temporal solitons in $\chi^{(2)}$ media in the case when the GVD (group-velocity dispersion) at the SH (second harmonic) is normal, which is the situation occurring in optical materials available to the experiment, and is a major issue impeding experimental studies of such solitons. A full range of essential parameters of the system was covered by the analysis, including the GVM (group-velocity mismatch) coefficient. The stability region for the solitons is limited by the onset of the fast growth of tails in the SH component (strictly speaking, some tail is always present in the solution if the GVD is normal at the SH, but the stability region can be defined so as to make the energy loss to the tail generation completely negligible under conditions relevant to any feasible experiment).

We have demonstrated that the stability region can be vastly expanded by means of two "management" techniques, *viz*., the DM (dispersion management), and GVM management. In the former case, the normal GVD at the SH is periodically compensated by layers of a material with anomalous dispersion. In the latter case, the solitons are readily stabilized by periodic mutual compensation of positive and negative GVM. A peculiarity of the DM scheme elaborated here for the $\chi^{(2)}$ media, in comparison with the well-known counterpart of this technique in $\chi^{(3)}$ media, is that the DM-stabilized $\chi^{(2)}$ solitons feature very weak intrinsic oscillations. In addition to the systematic numerical results, a crude analytical approximation for the DM was developed too, aimed at understanding the suppression of the tail formation under the action of this technique.

Experimental conditions which should allow the realization of the proposed management schemes may be similar to those under which the walkoff-compensating tandem scheme has been realized [23]. That work used a set of ten alternating plates cut (at different orientations) from a KTP crystal, with the thickness of each plate 1 mm. Following the same work, it is expected that picosecond beams with the transverse waist ~20 μm can be used to create the solitons.

**Acknowledgement**